\documentclass[aps,pra,reprint,superscriptaddress]{revtex4-1}
\usepackage{amsmath,amssymb,amsfonts}
\usepackage[english]{babel}
\usepackage{graphicx}

\begin{document}

\title{Exact dynamics of finite Glauber-Fock photonic lattices}

\author{B. M. Rodr\'{\i}guez-Lara}
\affiliation{Centre for Quantum Technologies, National University of Singapore, 2 Science Drive 3, Singapore 117542. }

\begin{abstract}
The dynamics of Glauber-Fock lattice of size N is given through exact diagonalization of the corresponding Hamiltonian; the spectra $\{ \lambda_{k} \}$ is given as the roots of the $N$-th Hermite polynomial, $H_{N}(\lambda_k/\sqrt{2})=0$, and the eigenstates are given in terms of Hermite polynomials evaluated at these roots. 
The exact dynamics is used to study coherent phenomena in discrete lattices. 
Due to the symmetry and spacing of the eigenvalues $\{ \lambda_{k} \}$, oscillatory behavior with highly localized spectra, that is, near complete revivals of the photon number and partial recovery of the initial state at given waveguides, is predicted.
\end{abstract}


\maketitle

\section{Introduction} \label{sec:S1}

Waveguide lattices, that is, arrays of single-mode waveguides coupled by evanescent fields, have been the focus of considerable interest due to their ability to simulate a variety of quantum effects under negligible decoherence.
Examples of such quantum effects are Bloch oscillations in lattices with linearly varying  on-site refraction index \cite{Peschel1998p1701,Pertsch1999p4752,Morandotti1999p4756,Rai2009p053849}, Zeno effect due to a defect in the coupling between the first two waveguide in a lattice with otherwise homogeneously coupled components \cite{Longhi2006p110402}, random walks in homogeneous lattices \cite{Rai2008p042304,Perets2008p170506,Bromberg2009p253904}, Anderson localization in homogeneous lattices where controlled disorder has been added \cite{Thompson2010p053805,Martin2011p13636} and the effect of isolated defects on quantum correlations \cite{Longhi2011p033821}.
In particular, lattices where the coupling is homogeneous are well understood and their analytical closed form time evolution is well known \cite{Jones1965p261,Makris2006p036616,Rai2008p042304,SotoEguibar2011p158}.

Recently, a photonic waveguide lattice where the coupling between adjacent waveguides varies as the square root of their position in the lattice has been proposed. The propagation of a classical field in a semi-infinite array of this type, the so-called Glauber-Fock photonic lattice, has been solved in close analytical form by creatively mapping the $j$-th waveguide to the $j$-th Fock state, this has been shown to produce a classical analogue to quantum coherent and displaced Fock states at the lattice output \cite{PerezLeija2010p2010}. 
The quantum correlations of non-classical light input have also been analyzed by numerical diagonalization and classical experimental results for single waveguide input have been presented for a lattice composed of sixty waveguides \cite{Keil2011p103601}. 

Here, in Section \ref{sec:S2}, it is shown that the finite Glauber-Fock Hamiltonian describing an array of identical waveguides where nearest neighbor couplings varies as the square root of the position of the waveguide is exact diagonalizable.
The exact result is given in terms of Hermite polynomials evaluated at the roots of the $N$-th Hermite polynomial, where $N$ is the number of waveguides in the system.
In Section \ref{sec:S3}, it is shown that, for a Fock state coupled to the zeroth waveguide, there exists an almost complete revival of the probability to find the photons back in the starting waveguide; the opposite occurs when the state couples to the end waveguide, where oscillations are polychromatic and revivals are weak.
Single-waveguide revivals do not occur in semi-infinite Glauber-Fock lattices, nor in uniform lattices unless multi-input phenomena or tunning of the lattice is used; say, Talbot effect \cite{Iwanow2005p053902} or Bloch oscillations \cite{Peschel1998p1701,Pertsch1999p4752}, in that order. 
Single input revivals occur in more complex waveguide lattices, for example, harmonic oscillator \cite{Gordon2004p2752}, Jaynes-Cummings \cite{Longhi2011} and Glauber-Fock oscillator \cite{PerezLeija2011p1109871} lattices.
The dynamics of initial states involving multiple waveguides is presented in Section \ref{sec:S4}. 
Time evolution of two-waveguide input, in particular product and NOON states, is explicitly discussed and revivals for fidelities, in the single-photon superposition case, and two-photon correlations, for the two-photon case, are shown. 
Finally, in Section \ref{sec:S5}, conclusions are presented.  

\begin{figure}
\includegraphics[width= 0.8 \linewidth]{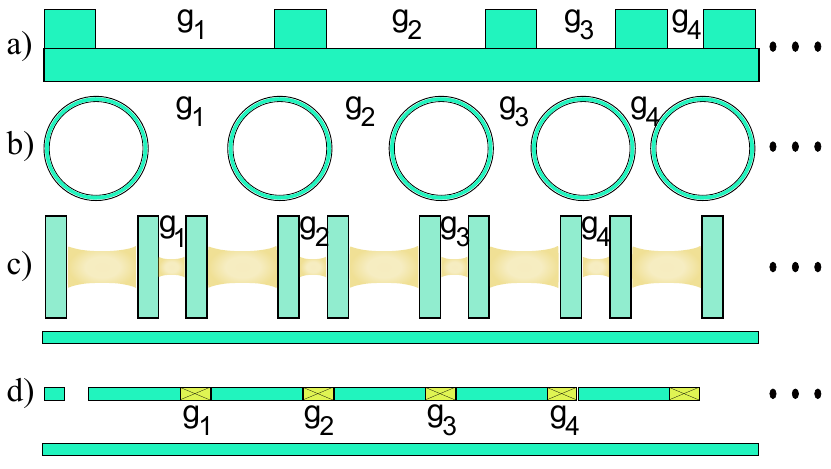} 
\caption {(Color online) Some systems modeled by the so-called Glauber-Fock Hamiltonian, Eq.\eqref{eq:GFH}.(a)
 Array of identical photonic waveguides, (b) micro-ring resonators, (c) coupled cavity arrays, and (d) capacitive coupled strip-line resonator arrays.}  \label{fig:Fig1}
\end{figure}

\section{Exact dynamics} \label{sec:S2}

The Hamiltonian describing a one-dimensional chain of $N$ cavities where nearest neighbors are coupled according to their position in the chain is given by, in units of $\hbar$,
\begin{eqnarray}
\hat{H} = \omega \sum_{j=0}^{N-1} \hat{a}^{\dagger}_{j} \hat{a}_{j} + g \sum_{j=0}^{N-2} \sqrt{j+1} \left( \hat{a}^{\dagger}_{j} \hat{a}_{j+1} + \hat{a}_{j} \hat{a}^{\dagger}_{j+1}\right). \label{eq:GFH}
\end{eqnarray}
The operator $\hat{a}^{\dagger}_{k}$ ($\hat{a}_{k}$) creates (annihilates) a photon in the $k$-th cavity, the constants $\omega$ and $g$ are the field frequency and the base coupling between cavities; these terms are related to the refraction index of the waveguides and the inter-waveguide distance in the photonic lattice. 
Figure \ref{fig:Fig1} shows a sampler of systems modeled by this Hamiltonian.

In the frame defined by the free field, $U_{0}(t) = e^{- \imath   \omega t \sum_{j=0}^{N-1} \hat{a}^{\dagger}_{j} \hat{a}_{j} }$, the dynamics is given by the Hamiltonian, in units of $\hbar g$,
\begin{eqnarray} \label{eq:HI}
\hat{H}_{I} =   \sum_{j=0}^{N-2} \sqrt{j+1} \left( \hat{a}^{\dagger}_{j} \hat{a}_{j+1} + \hat{a}_{j} \hat{a}^{\dagger}_{j+1}\right)
\end{eqnarray}
which Heisenberg equations of motion, $\dot{a}_{j} = \imath [\hat{H}_{I}, \hat{a}_{j}] $, are those of a Glauber-Fock photonic lattice \cite{PerezLeija2010p2010},
\begin{eqnarray}
 \imath \dot{a}_{0} &=&  \hat{a}_{1}, \\
 \imath \dot{a}_{j} &=&  \sqrt{j} \hat{a}_{j+1} + \sqrt{j-1} \hat{a}_{j-1}, \\
 \imath \dot{a}_{N-1} &=& \sqrt{N-2} \hat{a}_{N-2}. 
\end{eqnarray}
In matrix form, the Heisenberg set is given by the expression $ \partial_{t} \vec{a} =  -\imath \hat{M} \vec{a}$ where the matrix elements of $\hat{M}$ are $m_{j,k} =   \delta_{j+1,k} \sqrt{k} + \delta_{j,k+1} \sqrt{j}$, with  $j,k = 0,1,\ldots,N-1$. 
As the matrix $\hat{M}$ is real, symmetric and tridiagonal, there exists a solution for this differential set given by,
\begin{eqnarray}
\vec{a}(t) &=& e^{-\imath \hat{M} t} \vec{a}(0), \\
&=& \hat{V}^{\dagger} e^{-\imath \Hat{\Lambda} t} \hat{V} \vec{a}(0), \label{eq:HEMa}
\end{eqnarray}
where $\hat{\Lambda}$ is a diagonal matrix with elements $\lambda_{j}\equiv\lambda_{j,j}$ being the eigenvalues of $\hat{M}$, and the $j$-th eigenvector with $k$-th element $v_{jk}$ defines the $j$-th row of the matrix $\hat{V}$. As $v^{\ast}_{j,k} = v_{k,j}$, it is possible to write
\begin{eqnarray} \label{eq:afnt}
\hat{a}_{j}(t) &=& \sum_{k=0}^{N-1} U_{j,k}(t) \hat{a}_{k}(0), \\
\quad U_{j,k}(t) &=& \sum_{l=0}^{N-1} e^{- \imath \lambda_{l} t} v_{lj} v_{lk}.
\end{eqnarray}
The collective modes $\vec{b}= V \vec{a}$ allow rewriting the equation of motion for the annihilation operator, Eq.\eqref{eq:HEMa}, as
\begin{eqnarray}
\vec{b}(t) &=& e^{-\imath \Hat{\Lambda} t} \vec{b}(0), \end{eqnarray}
which leads to the diagonal Hamiltonian, in units of $\hbar g$,
\begin{eqnarray}
\hat{H}_{I} = \sum_{j=0}^{N-1} \lambda_{j} \hat{b}^{\dagger}_{j} \hat{b}_{j}.
\end{eqnarray}
Thus, the solution to Heisenberg equations of motion lead to the diagonalization of the Hamiltonian \cite{MarSarao2011p402}.

The method of minors \cite{Horn1990} delivers the eigenvalues $\lambda_{j}$ as the roots of the characteristic polynomial of the matrix $\hat{M}$ given by 
\begin{eqnarray}
H_{N}\left( \frac{\lambda_{j}}{\sqrt{2}} \right) = 0, \quad j=0,\ldots,N-1. 
\end{eqnarray}
where $H_{n}(x)$ is the $n$-th Hermite polynomial \cite{Lebedev1965} which zeros are well known \cite{Abramowitz1970}.
The eigenvectors are given by solving the eigenvalue equation $(M - I \lambda_{j}) \vec{v}_{j} = 0$ leading to the recurrence relations, 
\begin{eqnarray}
-\lambda_{j} v_{j,0} + \sqrt{1} v_{j,1} &=& 0, \\
\sqrt{k} v_{j,k-1} - \lambda_{j} v_{j,k} + \sqrt{k+1} v_{j,k+1} &=& 0, 
\end{eqnarray}
for $j = 0, \ldots, N-1$ and $ k=1,2,\ldots,N-2$. These recurrence relations are fulfilled by the matrix elements
\begin{eqnarray}
v_{j,k} &=& \frac{u_{j,k} }{\sqrt{ \sum_{k=0}^{N-1} u_{j,k}^2} }, \label{eq:vjk} \\
u_{j,k} &=& \frac{1}{\sqrt{2^{k} k!}} H_{k}\left( \frac{\lambda_{j}}{\sqrt{2}} \right),
\end{eqnarray}
for $j,k = 0, \ldots, N-1$. The Hamiltonian has been diagonalized by giving an analytical closed form.

\section{Single-waveguide input} \label{sec:S3}

Let us consider as initial state a Fock state with $m\ge1$ photons in the $p$-th waveguide,
\begin{equation}\label{eq:InFockS}
\vert \psi_{p}(0) \rangle = \frac{1}{\sqrt{m!}} \hat{a}^{\dagger m}_{p}(0) \vert 0 \rangle, \quad p=0,\ldots,N-1.
\end{equation}
The time evolution of the photon number at the $q$-th waveguide for such single input initial state is
\begin{equation} \label{eq:nktht}
\langle n_{q} \rangle_{p} \equiv \langle \psi_{p}(0) \vert \hat{a}^{\dagger}_{q}(t) \hat{a}_{q}(t) \vert \psi_{p}(0) \rangle = m \vert U_{p,q} (t) \vert^2.
\end{equation}

\begin{figure} 
\includegraphics[width= \linewidth]{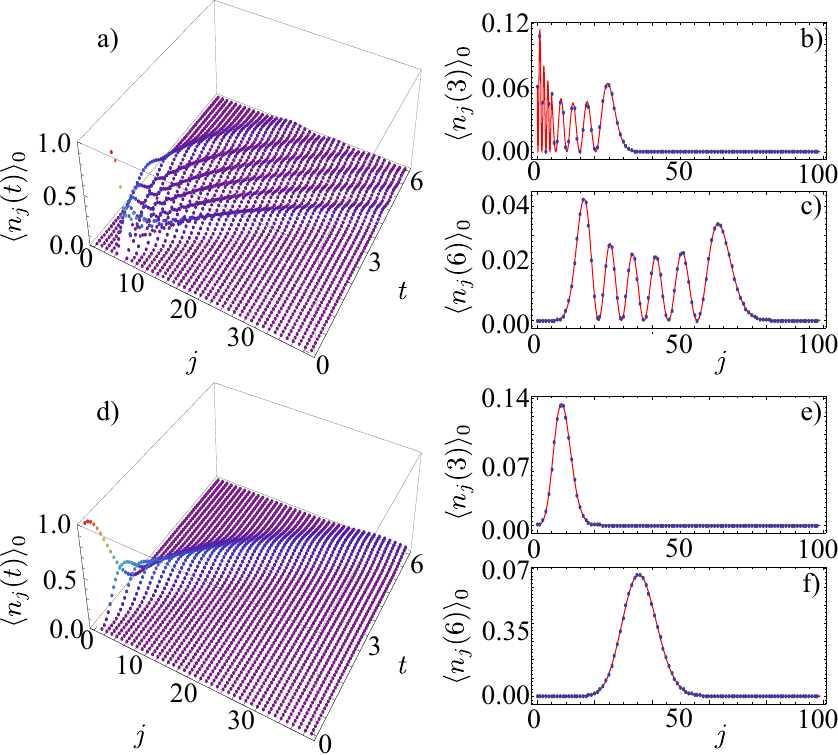} 
\caption{(Color online) Exact time evolution of the mean photon number at the $j$-th waveguide for an initial state consisting of a single-photon in the (a-c) fifth and (b-f) zeroth waveguide for a lattice size of two hundred waveguides. Snapshots at times (b,e) $t= 3$ and (c,f) $t=6$ comparing the exact finite result (blue dots) with the semi-infinite result (solid red line). 
Time is given in units of $g^{-1}$.} \label{fig:Fig2}
\end{figure}

For a single-photon coupled to a single-waveguide, it is possible to compare published results for a semi-infinite lattice \cite{PerezLeija2010p2010,Keil2011p103601} with the results presented here for a finite large lattice where propagation time is short enough to guarantee that the probability of finding the photon near the $(N-1)$-th waveguide is negligible. This leads to the following identities,
\begin{eqnarray}
\lim_{N \rightarrow \infty} \vert U_{k-s,n} \vert^2 &=& \vert e^{-t^2/2} (\imath t)^{s} \sqrt{\frac{(k-s)!}{k!}} L^{(s)}_{k-s}(t^2) \vert^2,  \label{eq:Conjecture1} \\
\lim_{N \rightarrow \infty} \vert U_{k+s,n} \vert^2 &=& \vert e^{-t^2/2} (\imath t)^{s} \sqrt{\frac{k!}{(k+s)!}} L^{(s)}_{k}(t^2) \vert^2, \label{eq:Conjecture2}
\end{eqnarray}
where the expression $L^{(\alpha)}_{j}(x)$ is a generalized Laguerre polynomial \cite{Lebedev1965}. In the case of the initial photon impinging the zeroth waveguide, the value becomes
\begin{eqnarray}
\lim_{N \rightarrow \infty} \vert U_{k-s,0} \vert^2 = \left\vert e^{-t^2/2}  \frac{t^{s}}{\sqrt{k!}} \right\vert^2. \label{eq:Conjecture3}
\end{eqnarray}
Figure \ref{fig:Fig2} shows the time evolution of the mean photon number for a lattice composed of two hundred waveguides.
The cases of a single-photon starting in the zeroth, Fig.\eqref{fig:Fig2}(a-c), and fifth, Fig.\eqref{fig:Fig2}(d-f), waveguide are presented; the figure is equivalent to those found in \cite{PerezLeija2010p2010, Keil2011p103601}.

\begin{figure} 
\includegraphics[width= \linewidth]{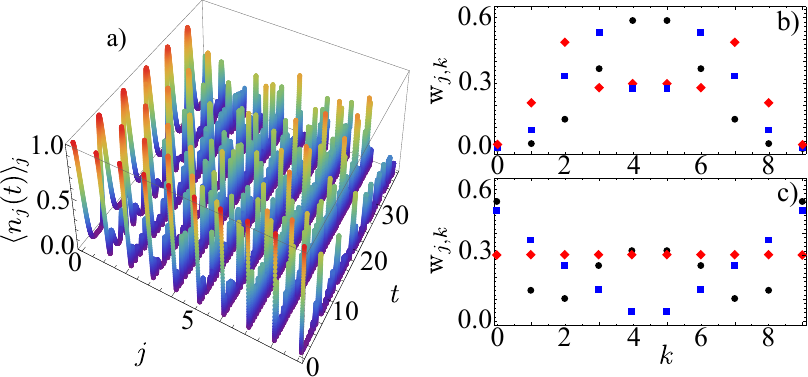} 
\caption{(Color online) Summary of the $N$ different cases where the initial state is given by a single-photon in the $j$-th waveguide. (a) Only the time evolution of the mean photon number at the $j$-th waveguide, $\langle n_{j}(t) \rangle_{j}$, is shown. Spectral decomposition $w_{j,k} = \vert\langle 0 \vert \hat{a}_j \hat{b}^{\dagger}_{k} \vert 0 \rangle \vert$ for the (b) first $j=1$ (black circles), $2$ (blue squares), $3$ (red diamond), and (c) last, $j=7$ (black circles), $8$ (blue squares), $9$ (red diamond), three waveguides.
Time is given in units of $g^{-1}$.} \label{fig:Fig3}
\end{figure}

Here, the focus will be the study of finite size behavior. Equation \eqref{eq:nktht} allows the emergence of well defined oscillations and almost complete revivals at the zeroth waveguide as $\langle n_{0} \rangle_{0} = m \vert U_{0,0} \vert^2$ with
\begin{eqnarray}
U_{0,0} = \sum_{l=0}^{N-1} \mathrm{w}_{0,l}  e^{-\imath \lambda_{l} t}, \quad \mathrm{w}_{0,l}= \frac{1}{\sum_{k=0}^{N-1} u_{l,k}^2}.
\end{eqnarray}
The weights $\mathrm{w}_{0,l}$ are highly centralized; that is, revivals at the zeroth waveguide are due to a few-component chromatic oscillation of the mean photon number; see Fig.\eqref{fig:Fig3}(b). 
Meanwhile, when the Fock state couples to the $(N-1)$-th waveguide, 
\begin{eqnarray}
U_{N-1,N-1} = v_{0,N-1}^2 \sum_{l=0}^{N-1}  e^{-\imath \lambda_{l} t}, \label{eq:Polychrome}
\end{eqnarray}
all eigenvalues contribute equally to the time evolution of the mean photon number $ \langle n_{N-1} \rangle_{N-1} = m \vert U_{N-1,N-1} \vert^2$ for any given size of the lattice and major revivals have a lower probability to appear than in the opposite extreme case, $\langle n_{0} \rangle_{0}$; see Fig.\eqref{fig:Fig3}(c). 

Figure \ref{fig:Fig3}(a) shows the time evolution of the mean photon number at the $j$-th waveguide for the $N$ possible initial conditions where a single-photon starts at the same waveguide, $\langle n_{j} \rangle_{j} = m \vert U_{j,j} \vert^2$. 
Figure \ref{fig:Fig3}(b) and (c) shows the spectral weights corresponding to the first and last three waveguides, in that order.
It is possible to see that major revivals of the mean photon number at the initial waveguide only occur when the input field impinges near the zeroth waveguide.

\begin{figure} 
\includegraphics[width= \linewidth]{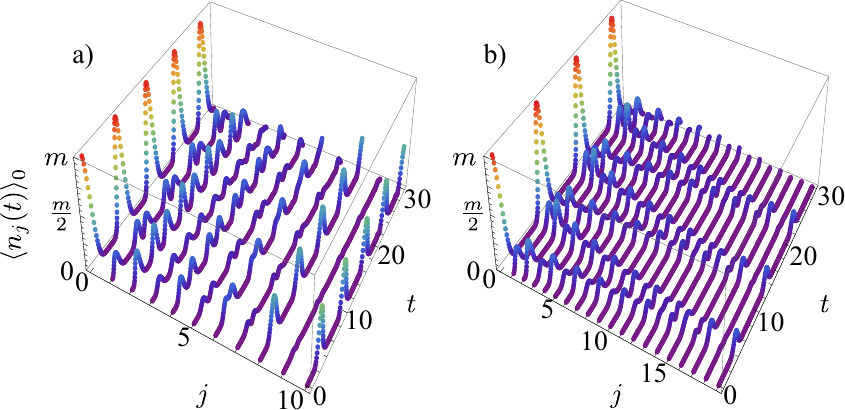} 
\caption{(Color online) Exact time evolution of the mean photon number at the $j$-th waveguide for an initial state consisting of a $m$-photon Fock state in the zeroth waveguide, $n_{0,j}$ from Eq.\eqref{eq:vjk}, for a lattice size of (a) $N=11$ and (b) $N=20$ waveguides. Time is given in units of $g^{-1}$.}
\label{fig:Fig4}
\end{figure}

Figure \ref{fig:Fig4} shows the time evolution of the mean photon number, Eq.\eqref{eq:nktht}, in a finite Glauber-Fock lattice composed by eleven, Fig\ref{fig:Fig4}(a), and twenty, Fig\ref{fig:Fig4}(b), waveguides for the initial state given in Eq.\eqref{eq:InFockS} at the zeroth waveguide, $p=0$. 
In this figure, it is possible to see an oscillator-like behavior for single-waveguide input with almost complete revivals of the mean photon number at the zeroth waveguide; well defined minor revivals occur at the last waveguide of the lattice too.

\section{Multi-waveguide input} \label{sec:S4}

The focus of this section will be the study of the oscillator-like behavior, already found for single-waveguide input, in the case of multiple-waveguide input. 
The dynamics for single-photon and a pair of relevant multi-photon, that is, product and NOON, states will be discussed.

\subsection{Single-photon superposition states}

\begin{figure} 
\includegraphics[width= \linewidth]{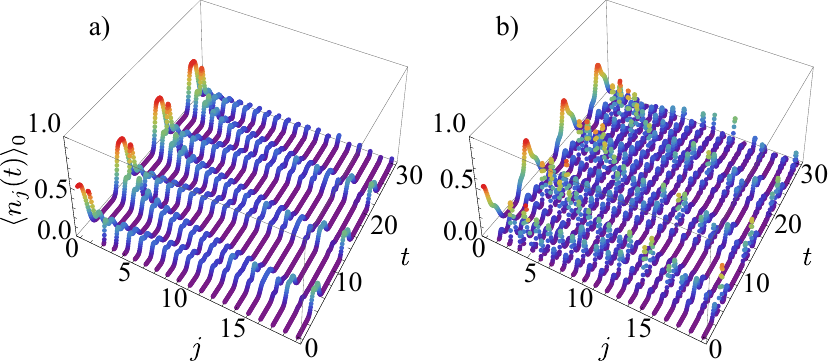} 
\caption{(Color online) Time evolution of the mean photon number, Eq.\eqref{eq:Belln}, for a
 Bell state, Eq.\eqref{eq:BellS} with $\alpha=\beta=1/\sqrt{2}$, coupled to the (a)$(j,k)=(0,1)$ and (b)$(j,k)=(0,4)$ waveguides. Time is given in units of $g^{-1}$.}
\label{fig:Fig5}
\end{figure}

The general state describing the superposition of one photon coupled to the lattice is 
\begin{eqnarray}
\vert \psi(0)_{s} \rangle = \sum_{j=0}^{N-1} c_{j} \hat{a}_{j}^{\dagger} \vert 0 \rangle, \quad \sum_{j=0}^{N-1} \vert {c}_{j} \vert^{2} = 1.
\end{eqnarray}
Such an initial state covers, for example, Bell states, when a single-photon is evenly coupled to two waveguides only, and W-states, when the photon is evenly coupled to all waveguides.
The mean photon number at the $q$-th waveguide evolution for such class of states is given by the expression, 
\begin{eqnarray}
\langle n_{q} \rangle_{ps} = \vert \sum_{j=0}^{N-1} c_{j} U_{j,q}(t) \vert^{2}.
\end{eqnarray}
The simplest of such states is the output of a beam splitter with transmission (reflection) $\alpha$ ($\beta= \sqrt{1 - \alpha^2}$) coupled to two waveguides,
\begin{eqnarray}
\vert \psi(0)_{s} \rangle &=& ( \alpha \hat{a}_{j}^{\dagger} + \beta \hat{a}_{k}^{\dagger} ) \vert 0 \rangle, \label{eq:BellS} \\ 
\langle n_{q} \rangle_{s} &=& \vert \alpha U_{j,q}(t) + \beta U_{k,q}(t) \vert^{2}. \label{eq:Belln}
\end{eqnarray} 
Figure \ref{fig:Fig5}(a) shows the evolution of a 50/50 beams splitter, that is, a Bell, initial state coupled to the $(j,k)=(0,1)$ waveguides and Fig. \ref{fig:Fig5}(b) for $(j,k)=(0,4)$. The oscillator-like behavior of the mean photon number evolution is well defined for the first set of initial conditions and noisy for the second. The revivals of the mean photon number at the starting waveguide are well defined and almost complete for both sets as shown in the following.

\begin{figure} 
\includegraphics[width= \linewidth]{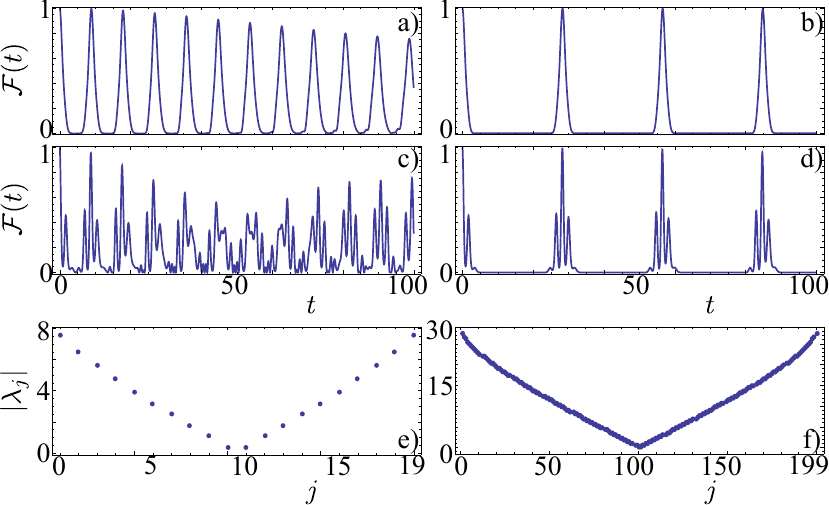} 
\caption{(Color online) Time evolution of the fidelity, Eq.\eqref{eq:BSFid}, coupled to the $(j,k)=(0,1)$ waveguides with (a) $N=20$ and (b) $N=200$ and to the $(j,k)=(0,4)$ waveguides with (c) $N=20$ and (d) $N=200$. Absolute value of the normal mode spectra for (e) $N=20$ and (f) $N=200$. Time is given in units of $g^{-1}$.}
\label{fig:Fig6}
\end{figure}

In order to quantify the likeliness between the initial state and its time evolution, it is possible to define a fidelity,
\begin{eqnarray}
\mathcal{F}(t) &=& \left\vert  \langle \psi(0)_{s} \vert \psi(t)_{s} \rangle \right\vert^2 \nonumber \\
&=&  \left\vert \vert \alpha \vert^2 U_{j,j} + \vert \beta \vert^2 U_{k,k} + 2 \mathrm{Re}(\alpha^{\ast} \beta) U_{j,k}   \right\vert^2, \label{eq:BSFid}
\end{eqnarray}
that is, the fidelity reaches a value of one if the evolved state and the initial state are identical.
Building upon the spectral decomposition ideas shown in Fig.\eqref{fig:Fig3}, it is possible to realize that, in order to get strong revivals, the initial state should be coupled to the zeroth waveguide and the coupled waveguides should be as close as possible.
A lattice of larger size will present stronger revivals of the fidelity because the dominant eigenvalues of the spectra of $U_{j,j}$ and $U_{k,k}$ will be closer to each other allowing them to interfere constructively; of course, period of the revivals will increase with the size of the lattice and, for real world systems, losses will accumulate.
Figure \ref{fig:Fig6}(a,c) shows the time evolution of the fidelity, for the cases mentioned above, Fig.\ref{fig:Fig5}, and Fig.\eqref{fig:Fig6}(b,d) for identical initial conditions but a lattice of size two hundred waveguides, $N=200$. 
Figure \ref{fig:Fig6}(e,f) show the absolute value of the normal mode spectra of the lattice, $\{ \lambda_{j} \}$, ordered such that $\lambda_{0} < \lambda_{1} < \ldots < \lambda_{N-1}$.

\subsection{Product states}
\begin{figure} 
\includegraphics[width= \linewidth]{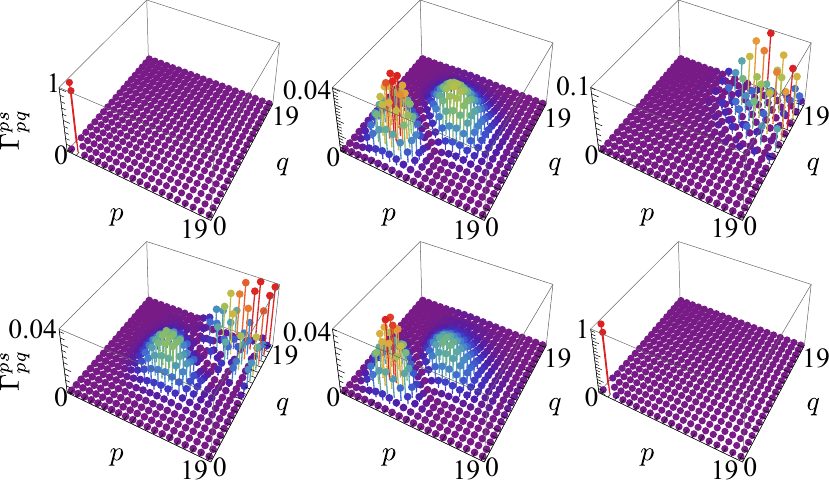} 
\caption{(Color online) Time evolution of the two photon correlation, Eq.\eqref{eq:2PhCorrProd}, for a two-photon product state, Eq.\eqref{eq:ProdS}, in a lattice of size $N=20$, with $(j,k)=(1,2)$ at times (a) $t = 0\times$, (b) $3\times$, (c) $5\times$, (d) $6\times$, (e) $7\times$, (f) $10\times \pi / (10 \vert\lambda_{10} \vert)$. Time in units of $g^{-1}$.}
\label{fig:Fig7}
\end{figure}

Published results for the semi-infinite Glauber-Fock lattice include the evolution of the two-photon separable state \cite{Keil2011p103601} belonging to the class,
\begin{eqnarray}
\vert \psi_{ps}(0) \rangle =  \prod_{j=1}^{k} \hat{a}^{\dagger}_{x_{k}} \vert 0 \rangle, 
\end{eqnarray}
where $\mathbf{x} = (x_{1},\ldots,x_{k})$ with $x_{i} \in [0,N-1]$ and $x_{i} \neq x_{j}$ for any $i \neq j$.
It is simple to calculate the evolution of the photon number at the $q$-th waveguide, 
\begin{eqnarray}
\langle n_{q} \rangle_{ps} = \sum_{j=1}^{k} \vert U_{x_{j},q} \vert^2.
\end{eqnarray}
That is, the probability of finding almost complete revivals in the oscillation of the mean photon number at the starting waveguides depends on the functions $U_{x_{j},q}$ having almost identical centralized spectral distributions; that is, the criteria commented above regarding distance to the zeroth waveguide, separation between field mode components, and size of the lattice hold.
For two-photon product states, the mean photon number evolution and the two-photon correlation function are well known \cite{Bromberg2009p253904,Keil2011p103601},
\begin{eqnarray}
\vert \psi(0)_{ps} \rangle &=& \hat{a}_{j}^{\dagger} \hat{a}_{k}^{\dagger} \vert 0 \rangle, \label{eq:ProdS} \\
\langle n_{q} \rangle_{ps} &=& \vert U_{j,q} \vert^2 + \vert U_{k,q} \vert^2 , \\
\Gamma_{pq}^{ps} &=& \vert U_{p,j} U_{q,k} + U_{p,k} U_{q,j} \vert^2 \label{eq:2PhCorrProd} .
\end{eqnarray}
The two-photon correlation can be used to verify the partial recovery of the initial state. 
Figure \ref{fig:Fig7} shows the time evolution of the two photon correlation for a two-photon product state in a lattice of size twenty waveguides. It is possible to see the partial recovery at time approximately equal to $t = \pi/\lambda_{\textrm{min}}$ as expected from all previous arguments; $\lambda_{\textrm{min}} = \vert \lambda_{N/2} \vert$ for both even and odd (with $N/2$ rounded to the next integer) lattice parameter $N$ and, again, $\lambda_{0} <  \lambda_{1} < \ldots < \lambda_{N-1}$.

\subsection{NOON states}
\begin{figure} 
\includegraphics[width= \linewidth]{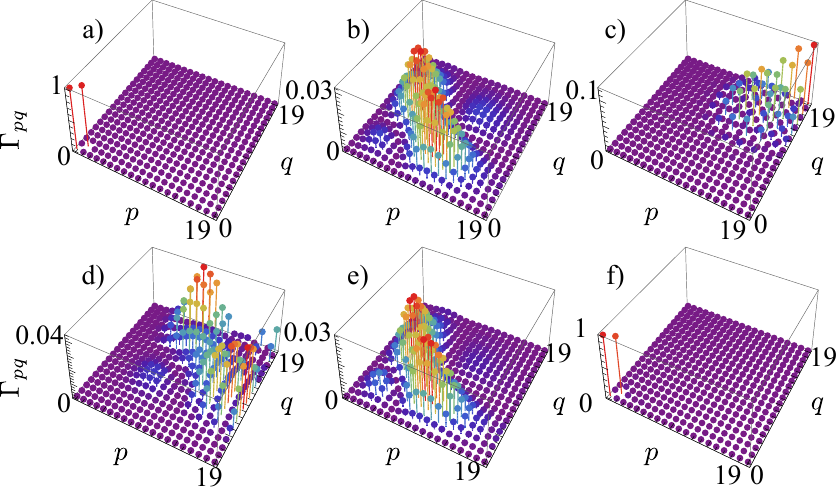} 
\caption{(Color online) Time evolution of the two photon correlation, Eq.\eqref{eq:2PhCorrNOON}, for a two-photon NOON state, Eq.\eqref{eq:NOONS} with $\phi = 0$, in a lattice of size $N=20$ with $(j,k)=(1,2)$ at times (a) $t = 0\times$, (b) $3\times$, (c) $5\times$, (d) $6\times$, (e) $7\times$, (f) $10\times \pi / (10  \vert\lambda_{10}\vert)$. Time in units of $g^{-1}$.}
\label{fig:Fig8}
\end{figure}

Higher order NOON states are a highly entangled class of states,
\begin{eqnarray}
\vert \psi(0) \rangle &=& \frac{1}{2 \sqrt{m!}} \left( \hat{a}_{j}^{\dagger m} + e^{ i m \phi} \hat{a}_{k}^{\dagger m} \right) \vert 0 \rangle, \quad m=2,3,\ldots  \label{eq:NOONS}\\
\langle n_{q} \rangle &=& \frac{m}{2} \left( \vert U_{j,q} \vert^2 + \vert U_{k,q} \vert^2  \right).
\end{eqnarray}
Notice that the restriction $m\ge2$ has been implemented to separate the single-photon superposition treated before which delivers an interferometer-like mean photon number evolution.
It is well known that for the two-photon case, $m=2$, the time evolution of the mean photon number is identical with that of the two-photon separable state considered above but the two-photon correlation is different \cite{Bromberg2009p253904}
\begin{eqnarray}
\Gamma_{pq} = \vert U_{p,j} U_{q,j}\vert^2 + \vert U_{p,k} U_{q,k} \vert^2 + 2 \mathrm{Re} \left( e^{i m \phi} U_{p,j}^{\ast} U_{q,j}^{\ast} U_{p,k} U_{q,k} \right). \label{eq:2PhCorrNOON}
\end{eqnarray}
Figure \ref{fig:Fig8} shows the time evolution of the two photon correlation for a two-photon NOON state in a lattice composed by twenty waveguides. Again, it is possible to see the partial recovery at time approximately equal to $t = \pi/\lambda_{\textrm{min}}$ as defined above.

\section{Conclusions} \label{sec:S5}

The exact dynamics of a Hamiltonian describing a finite array of coupled identical photonic waveguides where the coupling varies as the square root of the position of the waveguide in the lattice was presented. 

It was shown that the closed form analytical time evolution predicts a strong oscillator-like behavior of the lattice for single and double waveguide input. 
That is, the initial state is partially reconstructed periodically, leading to revivals of the mean photon number evolution at the waveguides where the photons started.
The strength of the reconstructions, thus of the revivals, is a combination of input distance from the zeroth waveguide, inter-waveguide input distance, and lattice size.

\begin{acknowledgments}
The author is grateful to Changsuk Noh and Rafael Rabelo for helpful discussion and acknowledges constructive criticism by Dimitris G. Angelakis and Amit Rai.
\end{acknowledgments}


\end{document}